\documentstyle[prl,aps,multicol,
epsf]{revtex}

\author{
S.N. Dorogovtsev$^{1, 2,}$\cite{S} and J.F.F. Mendes$^{1,}$\cite{F}
}
\title{
Time of avalanche mixing of granular materials in a half filled rotated drum
}
\address{
$^{1}$ Departamento de F\'\i sica and Centro de F\'\i sica do Porto, Faculdade de Ci\^encias, 
Universidade do Porto\\
Rua do Campo Alegre 687, 4169-007 Porto, Portugal\\
$^{2}$ A.F.Ioffe Physico-Technical Institute, 194021 St.Petersburg, Russia
}

\begin{document}

\maketitle
\begin{abstract}
The avalanche mixing of granular solids in a slowly rotated 2D upright drum is studied. 
We demonstrate that the account of the difference $\delta$ between the angle of marginal stability 
and the angle of repose of the granular material leads to a restricted value 
of the mixing time $\tau$ for a half filled drum. 
The process of mixing is described by a linear discrete difference equation.
We show that the mixing looks like linear diffusion of fractions with the
diffusion coefficient vanishing when $\delta$ is an integer part of $\pi$. 
Introduction of fluctuations of $\delta$ supresses the singularities of $\tau(\delta)$ 
and smoothes the dependence $\tau(\delta)$.
\end{abstract}


\begin{multicols}{2}

\narrowtext


The evolution of granular materials in rotated drums is one of the most interesting topics of soft matter physics \cite{otbook,nedbook,durbook,r00,po,ris1,raj,baum,zik,clem,can2}. Granular flows induced in such a way demonstrate a lot of intriguing phenomena including, e.g., segregation of different fractions, mixing, self-organization, self-organized criticality, etc \cite{metcal,sam99,y99,ljc97,hk94,bw96,mcs97,m99,ev99}. In principle, the system is three-dimensional, and the process features depend strongly on the particular characteristics of the grains involved. 
A theoretical description of these processes is still unsolved problem, and only the simplest cases were considered analytically. 

Decrease of the dimensionality of the system by one simplifies the consideration, 
although does not make the arising problems to be less interesting. The avalanche mixing phenomenon \cite{metcal} observed in a slowly rotated 2D upright drum, i.e., in a flat disc, turned to be so impressive that one of figures of the paper has made a front page of the {\em Nature}. 

\begin{figure}[\!h]
\epsfxsize=3.5in
\epsffile{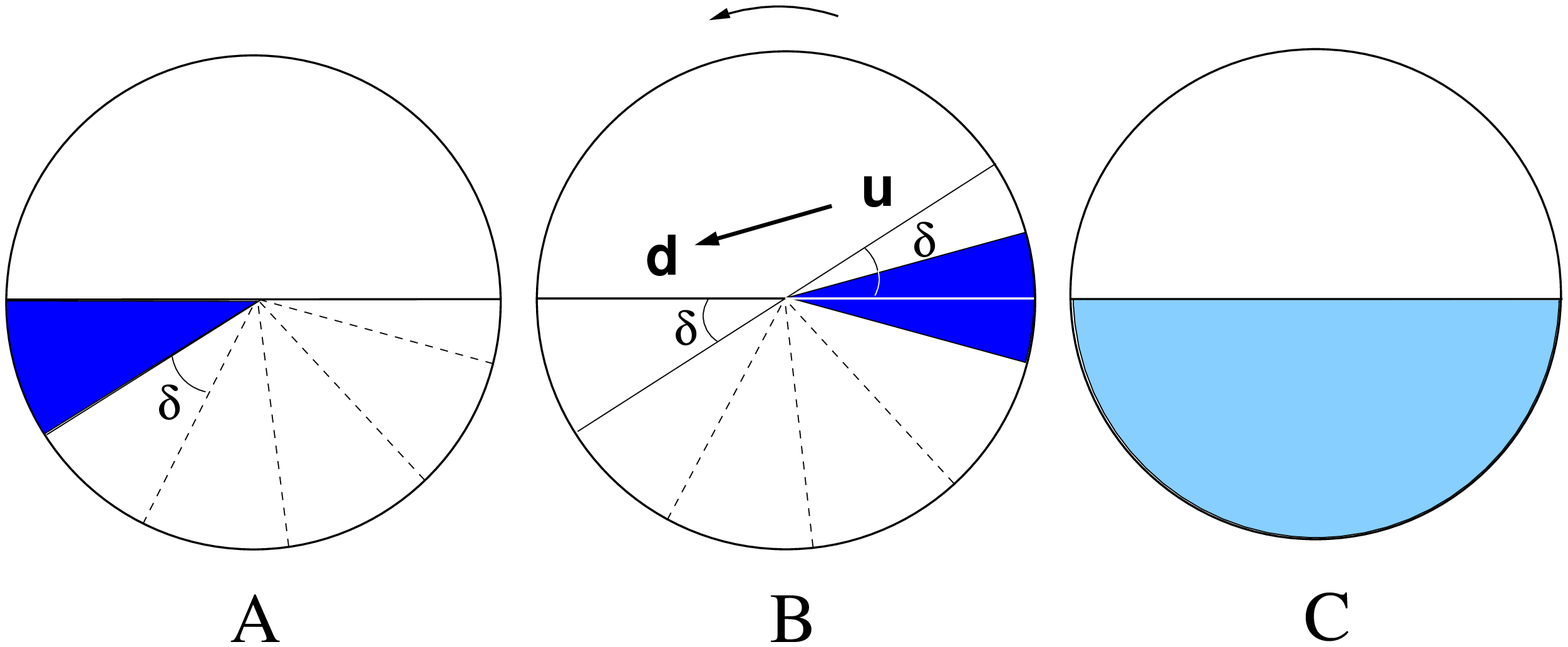}
\caption{
Schematic illustration of the avalanche mixing in a half filled 2D drum. 
Mixing begins from an inhomogeneous distribution of the fractions, 
e.g., from the configuration (A). (B) The drum starts rotate anticlockwise. 
When the free surface slope
attains the marginal stability angle value, the grains
of different fractions flow quickly from sector {\bf u} to sector {\bf d},
undergoing mixing. Then the free surface slope turns to be equal to
the repose angle of the material.
The angle $\delta$ is the difference between the angle of marginal
stability and the angle of repose. In the figure, the angle of repose is zero. 
If $\delta$ is not an integer part of $\pi$ or it fluctuates with time, finally 
the material is homogeneously mixed (C).
\label{fig1}}
\end{figure}

Let us introduce the avalanche mixing problem 
\cite{metcal,dor1,peratt1,kmso97,dor2,peratt2}.
A partially filled 2D upright drum rotates very slowly around its axis (see Fig.~\ref{fig1}). 
Small grains of the granular material move as a whole with the drum. 
There is, however,  
one exception: periodically, when the grains appear on the free surface, they 
fall down along it. If, initially, there is several separated fractions of the material, 
the grains of the different fractions may mix only in the avalanches 
which fall down along the free surface. One has to describe the  
evolution of the mixing, i.e., to describe how the process develops. 
The main quantity under study is the time of mixing $\tau$, that is the characteristic
time after that the material will be mixed homogeneously. It may be introduced, since the relaxation to the final homogeneous state in such a process proceeds by an exponentional law.

There are two particular values of the angle of the free surface slope of a granular material -- 
the angle of marginal stability and the angle of repose.  
(We assume that the free surface is straight. That is reasonable for a slowly rotated drum.) 
When the free surface slope achieves the angle of marginal stability, the grains  
from the sector of the angle $\delta$ near the upper half of
a free surface fall down (see Fig.~\ref{fig1} -- grains fall down
from sector {\bf u} to sector {\bf d}). 
Here, $\delta$ is the difference
between the angle of marginal stability and the angle of repose. 
The avalanche stops when the free surface slope approaches the repose angle, 
and the drum has to turn by the angle $\delta$ 
to 
launche the next avalanche. 
Therefore, the only reason for mixing in the problem under consideration is the granular
flow (avalanches) between these wedges. 

We assume that the fractions become to be mixed homogeneously after
the fall of the avalanche -- the mixture in sector {\bf d} in Fig.~\ref{fig1}. This case,
in which the avalanche mixing proceeds most quickly,
appears to be near the experimentaly realized situation \cite{metcal}. 
At least, using this assumption, one may obtain the minimal possible time of the avalanche mixing.

When $\delta \to 0$, the mixing was described in the frames of a simple
geometrical approach \cite{metcal,dor1,peratt1,dor2,peratt2}, and
an analytical theory \cite{dor1,dor2} describing the experimental observations \cite{metcal} surprisingly well has been proposed. 
Nevertheless, there is a very interesting particular case of the half filling, 
in which one certainly can not assume that $\delta$ is zero.  
(Note, that in the experiment of Metcalfe et al \cite{metcal}, 
$\delta \approx 8^\circ$.)  
Indeed, this 
assumption leads immediately to an infinite value of the mixing time for 
a half filled drum: there is no any mixing between the grains of the adjacent wedges 
if $\delta=0$, and the material will never mix. 

The finite difference between the angle of marginal stability and the repose angle 
makes $\tau$ to be restricted even if the drum is half filled. Hence, the 
finiteness of  
$\delta$ determines the avalanche mixing process in this particular case that is considered 
here.
   
Let us assume that there are two fractions -- black and white, and that the small grains 
of the different fractions are distinguished from each other only by their color. 
The material in sector {\bf d} of the angle $\delta$ is mixed homogeneously after the avalanches falls. 
Then, after the first half of turn, 
the mixing dynamics can be described by a set of concentration values, $\{c_m\}$, where $c_m$ is
the concentration of the black fraction in sector $\bf d$ (Fig.~\ref{fig1})
after the $m$-th avalanche falls, that happens at time
$t=m\delta$. Here, time is an angle of drum rotation. 
Note that a drum radius is not essential for the consideration 
and does not appear in our expressions.  
In the case of a half filled drum, the equation
for $c_m$ turns to be very simple:

\begin{equation}
\label{L1}
c_m = 
\left\{1\!-\! \left(\frac\pi\delta\!-\!\left[\frac\pi\delta\right]\right) \right\}
c_{m-[\pi/\delta] }+
\left\{\frac\pi\delta
\!-\! \left[\frac\pi\delta\right] \right\} c_{m-[\pi/\delta]-1}  .
\end{equation}
It can be understood easily from Fig.~\ref{fig1}.  
The grains of only two adjacent wedges mix. In each of these wedges the material is homogeneous after the previous avalanches, 
so two terms in the right part of Eq.~(\ref{L1}) are the contributions of the black 
grains from one of these two wedges, taken with the corresponding weight -- their relative angles 
in sector {\bf u} in Fig.~\ref{fig1}. 
Here, $[\ ]$ denotes an integer part of a number.
The values $c_m, \, m=-[\pi/\delta],-[\pi/\delta]+1,\ldots,-1,0$
may be taken as an initial condition if one does not consider the first half of turn. 
Obviously, the initial conditions are not essential for the mixing time value. 

For brevity, we use the notation
$n \equiv [\pi/\delta]$ for the integer part and
$a \equiv \pi/\delta-[\pi/\delta]$ for the fractional part.
With these notations, Eq. (\ref{L1}) may be written in a more compact form:

\begin{equation}
\label{2}
c_{m+n} = c_{m} - a(c_m-c_{m-1}) \,  .
\end{equation}

\begin{figure}[\!h]
\epsfxsize=3.1in
\epsffile{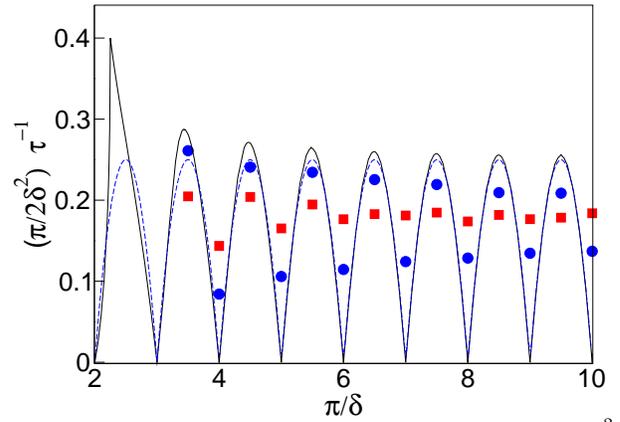 }
\caption{
\narrowtext
Scaled inverse avalanche mixing time $(\pi/2\overline{\delta}^2)\tau^{-1}$ vs 
$\pi/\overline{\delta}$. 
The solid line describing the case without fluctuations of $\delta$ is obtained from the solution of the characteristic
polynomial, Eq. (\protect\ref{3}).
The dashed line represents Eq. (\protect\ref{8})
which is, strictly speaking, found for small $\delta \ll \pi$. 
The points are results of the direct simulation of the evolution of the mixing with account for the fluctuations of $\delta$, Eq. (\protect\ref{10}) was used. The squares correspond to the homogeneous 
distribution of $\delta$ between $(1-1/8)\overline{\delta}$ and (1+1/8)$\overline{\delta}$, 
the circles -- to the homogeneous distribution of $\delta$ between $(1-1/4)\overline{\delta}$ and (1+1/4)$\overline{\delta}$.   
The lines between the points are not shown.  
\label{fig2}}
\end{figure}

According to usual prescriptions \cite{pinney},
one may search for the solution of Eq.~(\ref{L1}) or Eq.~(\ref{2})
in the form of
a linear combination of the terms $\lambda_j^m$, where $\lambda_j$
are the roots of the characteristic polynomial

\begin{equation}
\label{3}
\lambda^{n+1}-(1-a)\lambda-a=0 \ .
\end{equation}
To obtain the long time relaxation we should
find, among the roots $\lambda_j$ of Eq.~(\ref{3}), the maximal value $|\lambda_j| \equiv \lambda_{max}$ that is lower then the
root $\lambda=1$ 
(note that there is no multiple roots; 
$\lambda=1$ corresponds to a stationary situation). 
The mixing time $\tau$ is expressed in the
terms of $\lambda_{max}$:
$\tau^{-1}=(1/\delta)\log(1/\lambda_{max})$,
so it may be calculated directly for any value of $a$ and a finite $n$.
The result is shown by the solid line in Fig.~\ref{fig2}. One can see 
that the mixing time is infinite for only integer values of $\pi/\delta$. It has local minima near 
the points $\pi/\delta=k+1/2$, where $k=2,3,\ldots$.

The process of the mixing may be interpreted in the following way. It follows from Eq. (\ref{2}) that the long scale and the short one are separated if $n \ll 1$. Therefore, in such a situation, it is reasonable to use different variables for these two scales. Let us introduce a new time variable $\tilde{t}$. We observe periodically the system under study with the interval of periodicity $\pi$ (e.g., using flashes like in the stroboscopic effect) and describe the result of the observation by the quantity $c(\varphi,\tilde{t})$, the concentration of the black fraction at the angle $\varphi$ at the ``time'' $\tilde{t}$. Here, $0<\varphi<\pi$. 

At $n \ll 1$, in the continuous limit, $c_m \to c(m\delta)$. Subtracting $c((m-a)\delta)$ from both sides of Eq. (\ref{2}) and using the obvious relation, $(n+a)\delta = \pi$, we obtain immediately the equation for $c(\varphi,\tilde{t})$:

\begin{equation}
\label{5}
\frac{\partial c(\varphi,\tilde{t})}{\partial\tilde{t}}=
\frac{\delta^2 a(1-a)}{2\pi}\,
\frac{\partial^2 c(\varphi,\tilde{t})}{\partial\varphi^2} \, . 
\end{equation}
Hence, in such an approach, the mixing looks like the diffusion governed by Eq. (\ref{5}). We do not write out its solution since only the mixing time, i.e., the time of the exponentional relaxation to a homogeneous state, is interesting for us.  It follows from the diffusion coefficient and is of the form:

\begin{equation}
\label{8}
\tau^{-1}= \frac{2}{\pi} \delta^2 \left(\frac\pi\delta-
\left[\frac\pi\delta\right]\right)
\left\{1- \left(\frac\pi\delta-\left[\frac\pi\delta\right]\right) \right\} \, .
\end{equation}
This result (which is obtained in the limit $\delta \ll \pi$) is shown by the dashed curve in Fig. 2. One may see that the deviations from the results obtained directly from the characteristic polinomial equation are small even for large $\delta$. 
The considerable difference is visible only in the unphysical range, $\pi/3 < \delta < \pi/2$. 

Above, we considered the fixed $\delta$. Nevertheless, really, 
in an experiment, the angle $\delta$ is is not fixed but fluctuates with time. 
A particular form of the distribution function of $\delta$ is dependent on characteristics 
of the mixed grains and is not discussed here. 
What is the effect of these fluctuations?

Let the difference $\delta_m$ between the marginal stability angle and the repose angle for the $m$-th avalanche be 
fluctuating and, hence, depending on $m$. Let the distribution function for it be $P(\delta)$. 
One introduces the total angle of the drum turn, $ \theta_m = \sum_{i=1}^m \delta_i $. Then, after the $m$-th avalanche, the wedge between $\theta_{m-1}$ and $\theta_m$ of width $\delta_m$ is homogeneously mixed with the concentration of the black fraction $c_m$. 

The material for the mixing is taken from the wedge  
$\theta_{m-1} - \pi < t < \theta_m - \pi$. 
Now, unlike the previously considered case without the fluctuations, it may be more than two wedges with different concentration of the black grains in this sector. 
Therefore, if we assume 
$\theta_{m-1} - \pi < \theta_j<\ldots<\theta_{j+k} < \theta_m - \pi$, where the value of $k$ depends on the particular set of $\{\delta_m\}$, one obtaines the following equation:

\begin{eqnarray}
\label{10}
c_{m} = & & \frac{1}{\delta_m} \{
[\theta_j - (\theta_{m-1}-\pi)]c_j + \delta_{j+1}c_{j+1} + \ldots + 
\nonumber
\\[1ex]
& & \delta_{j+k}c_{j+k} +
[(\theta_{m}-\pi) - \theta_{j+k}]c_{j+k+1}   
\} \,  
\end{eqnarray}
that is the strict generalization of Eq. (\ref{L1}).
We study this equation numerically with account of the fluctuations of $\delta_m$ to find the mixing time. The results for the homogeneous distributions 
$P(\delta) = \Theta(\delta-(1-\Delta)\overline{\delta}) \Theta((1+\Delta)\overline{\delta}-\delta)/(2\Delta)$ 
with $\Delta=1/8$ and  $\Delta=1/4$ are presented in Fig. \ref{fig2} (here, $\Theta(\ )$ is the theta-function). 
(Note, that, in Fig. \ref{fig2}, we do not connect the points obtained from the numerics.)
One sees that the singularities of $\tau$ are effectively suppressed by the fluctuations. 
The oscillations of the dependence decrease with decrease of $\delta$, 
and $(\pi/(2\delta^2))\tau^{-1}$ approaches a constant value at small $\delta$. 
For small values of the parameter $\Delta$, 
this value is approached more quickly than for large ones but it is easy to check that, 
for all $C>0$, it equals approximately $0.175$. That is close to $1/6=\int_0^1 da\, a(1-a)$, 
i.e., to the average value of $(\pi/(2\delta^2))\tau^{-1}$ at small $\delta$. 

In the experiment \cite{metcal},
$\delta=8^\circ \pm 2^\circ$. From our estimation, that leads to $\tau \sim 70$ turns
of the drum. There are no available experimental data yet to be compared with this value. 
The measuring of $\tau$ in such a situation would be really interesting. 
One can expect that our 
estimation is quite reasonable because the agreement of the results of the geometrical 
approach \cite{dor2} for $\delta=0$ with the experiment \cite{metcal} is excellent.

In conclusion, we have described analytically the dynamics of the avalanche
mixing in the case of a half filled slowly rotated drum when the difference $\delta$ between
the angle of the marginal stability and the angle of repose plays the
principal role determining the value of the mixing time. 
The mixing looks like diffusion of grains between wedges with the diffusion
coefficient vanishing when the angle $\delta$ is an integer part of $\pi$. In these points, 
the dependence $\tau(\delta)$ has singularities which may be effectively suppressed by 
fluctuations of $\delta$. The proposed approach is, in fact, geometrical, and does not 
contain any other parameters apart of $\delta$ or its distribution. 
Two main assumptions were made: 
(i) We considered the granular material consisting of very small grains (much smaller than 
the drum size). 
(ii) The mixing of the fractions after each elementary avalanche were 
assumed to be homogeneous. 
These assumptions reserve space for future study of the avalanche mixing problem.

SND thanks PRAXIS XXI (Portugal) for a research grant PRAXIS XXI/BCC/16418/98. 
JFFM was partially supported by the projects PRAXIS/2/2.1/FIS/299/94. We also thank 
E. Lage for reading the manuscript and A.N.~Samukhin and H.~Watanabe for many useful discussions.\\


\vspace{3cm}
\vspace{.3in}

\end{multicols}
\end{document}